\documentclass[twocolumn,           % Format:   preprint, twocolumn
               showpacs,            % Pacs  :   showpacs, noshowpacs
               preprintnumbers,     % Preprint: preprintnumbers, 	
               aps,                 % Society:
               prd,          	    % Style : pra,prb,prc,prd,pre,prl,
               a4paper,             % Size  :   a4paper, ...
               groupedaddress,  % Title :   groupedaddress, 
               unsortedaddress,
               nofootinbib,         % Foonote:  footinbib, nofootinbib 
               tightenlines,        % Remove additional spaces in a line
               floats               % Floating pictures and tables
               ]{revtex4}

\usepackage{graphicx}
\usepackage{amssymb,latexsym}
\usepackage[draft=false]{hyperref}

\newcommand{\ba}{\begin{eqnarray}}
\newcommand{\ea}{\end{eqnarray}}  
\newcommand{\be}{\begin{equation}}
\newcommand{\ee}{\end{equation}}

\begin{document}

\hyphenation{brane-world}  

%%--- DRAFTCOPY --------------------------------
%% Prints a large "DRAFT" diagonally across each page
%% Does not show up in TeXview
%% \typeout{Prints "DRAFT" on each page; does not show in TeXView}
% \special{!userdict begin /bop-hook{gsave 200 30 translate
% 65 rotate /Times-Roman findfont 216 scalefont setfont
% 0 0 moveto 0.90 setgray (DRAFT) show grestore}def end}
%%------------------------------------------------

%======================================%
%<<<<<<<<<<<< TITLE PAGE >>>>>>>>>>>>>>%
%======================================%

%\renewcommand{\topfraction}{0.99}
%\renewcommand{\bottomfraction}{0.99}

\title{On the inflationary flow equations}
\author{Andrew R.~Liddle}
\affiliation{Astronomy Centre, University of Sussex, 
             Brighton BN1 9QJ, United 
Kingdom}
\date{\today} 
\pacs{98.80.Cq \hfill astro-ph/0307286}
\preprint{astro-ph/0307286}

%======================================%
%<<<<<<<<<<<<< ABSTRACT >>>>>>>>>>>>>>>%
%======================================%

\begin{abstract}
I explore properties of the inflationary flow equations. I show that the flow 
equations do {\em not} correspond directly to inflationary dynamics. 
Nevertheless, they can be used as a rather complicated algorithm for generating 
inflationary models. I demonstrate that the flow equations can be solved 
analytically and give a closed form solution for the potentials to which flow 
equation solutions correspond. I end by considering some simpler algorithms for 
generating stochastic sets of slow-roll inflationary models for confrontation 
with observational data.
\end{abstract}

\maketitle
%======================================%
%<<<<<<<<<<<< MAIN TEXT  >>>>>>>>>>>>>>%
%======================================%

%%%%%%%%%%%%%%%%%%%%%%%%%%%%%%%%%%%%%%%%%%%%%%%%%%%%%%%%%%%%%%%%
\section{Introduction}

The inflationary flow equations were introduced by Hoffman and Turner \cite{HT}, 
and have been proposed by Kinney \cite{K} as a way of generating large numbers 
of slow-roll inflation models which can be compared to observational data. They 
rely on defining a set of functions, the slow-roll parameters, based on 
derivatives of the Hubble parameter during inflation, and deriving a set of 
equations for their variation with time. Integration of these equations yields a 
trajectory in slow-roll parameter space, which can be interpretted as a  
variation with scale of the scalar and tensor spectra, usually written in terms 
of quantities such as the tensor-to-scalar ratio and the perturbation spectral 
indices and their running.

Although these equations have been employed to make comparisons with 
observations \cite{HK,peiris,KKMR}, as yet no clear connection has been made 
between the inflationary dynamics and the flow equations. As I will explain in 
this paper, there is in fact {\em no} direct connection between these two; the 
flow equations do not encode any physical model of inflationary dynamics. 
Despite this, it turns out 
that they can be used to generate inflationary models, though they represent 
quite a complicated algorithm for doing so. As it happens, this can be 
highlighted for slow-roll inflation by obtaining a closed form analytical 
solution to the flow equations and their relation to the inflationary potential. 

\section{The flow equations and their relation to inflation}

\subsection{The flow equations}

In this section I follow closely the notation and presentation by Kinney 
\cite{K}. We take as our fundamental quantity the evolution of the Hubble 
parameter $H$ as a function of $\phi$ (often called the Hamilton--Jacobi 
approach to inflation \cite{SB}). From this we define a set of Hubble slow-roll 
parameters,\footnote{This approach was originally suggested by Liddle, Parsons 
and Barrow \cite{LPB}, but here I follow the notation of Kinney \cite{K} rather 
than that paper.}
\begin{eqnarray}
\label{e:eps}
\epsilon(\phi) & \equiv &  \frac{m_{{\rm Pl}}^2}{4\pi} \left( 
\frac{H'(\phi)}{H(\phi)} \right)^2 \,; \\
\label{e:srdef}
^{\ell}\lambda_{{\rm H}} & \equiv & \left( \frac{m_{{\rm Pl}}^2}{4\pi}
	\right)^\ell \, \frac{(H')^{\ell-1}}{H^\ell} \, 
	\frac{d^{(\ell +1)} H}{d\phi^{(\ell +1)}} \quad ; \quad \ell \ge 1
\end{eqnarray}
where primes are derivatives with respect to the scalar field. For example, the 
parameter $^1\lambda_{{\rm H}}$ equals $(m_{{\rm Pl}}^2/4\pi) H''/H$ and is 
often denoted $\eta(\phi)$.

As the successive parameters feature an ever high number of derivatives, we can 
construct a hierarchy of flow equations, with the derivative of $\epsilon$ given 
in terms of parameters up to $^1\lambda_{{\rm H}}$, the derivative of 
$^1\lambda_{{\rm H}}$ in terms of parameters up to $^2\lambda_{{\rm H}}$ etc. 
Rather than the derivative with respect to $\phi$, it is convenient to take the 
derivative with respect to the number of $e$-foldings of inflation $N$, using 
the relation
\begin{equation}
\frac{d}{dN} = \frac{m_{{\rm Pl}}}{2\sqrt{\pi}} \, \sqrt{\epsilon} \, 
\frac{d}{d\phi} \,.
\end{equation}
Kinney derives the flow equations, using a convenient definition $\sigma \equiv 
2(^1\lambda_{{\rm H}})-4\epsilon$, as
\begin{eqnarray}
\label{e:flow}
\frac{d\epsilon}{dN} & = & \epsilon(\sigma+2\epsilon) \,; \nonumber \\
\frac{d\sigma}{dN} & = & -5\epsilon\sigma - 12 \epsilon^2 +
	2(^2\lambda_{{\rm H}}) \,; \\
\frac{d(^\ell\lambda_{{\rm H}})}{dN} & = & \left[\frac{\ell-1}{2} \, \sigma
	+(\ell-2)\epsilon \right](^\ell\lambda_{{\rm H}})+^{\ell+1}\!\!\!
	\lambda_{{\rm H}} \; ; \; \ell \ge2 \,. \nonumber
\end{eqnarray}
 
In order to solve this infinite series, it must be truncated by setting a 
sufficiently high slow-roll parameter to zero, i.e.~$^{M+1}\lambda_{{\rm H}} = 
0$ for some suitably large $M$ such as $M=5$ \cite{K}. Eqs.~(\ref{e:flow}) then 
comprise a closed set, and can be integrated by choosing initial conditions for 
the parameters $\epsilon$, $^1\lambda_{{\rm H}}$, ..., $^M\lambda_{{\rm H}}$. 
Typically either at some point $\epsilon$ reaches one, indicating the end of 
inflation, or the model reaches a late-time attractor with perpetual inflation.

\subsection{Interpretation of the flow equations}

The main purpose of this paper concerns understanding the dynamical properties 
of the flow equations and how they relate to the inflationary dynamics. The 
answer is that they do not at all, because {\em the above discussion has been 
carried out without ever mentioning the main dynamical equation of inflation}. 
The 
missing equation is the Hamilton--Jacobi equation (equivalent to the Friedmann 
equation)
\begin{equation}
\label{e:HJ}
[H'(\phi)]^2 - \frac{12\pi}{m_{{\rm Pl}}^2} H^2(\phi) = -\frac{32\pi^2}{m_{{\rm 
Pl}}^4} V(\phi)\,,
\end{equation}
which tells us how the expansion rate is linked to the potential $V(\phi)$ for 
the inflaton, which has also not been mentioned at all up to this point.

In fact, had the flow equations been written using $d/d\phi$, they would have 
amounted to a trivial set of relations amongst derivatives of the Hubble 
parameter; substituting in the definitions in terms of $H(\phi)$ would lead to a 
set of tautologous equations. The situation is made less trivial by the use of 
$N$ rather than $\phi$, which does require some dynamical information from the 
equations of motion; however either of these parameters is just measuring the 
distance along the trajectory, and does nothing to alter the actual trajectories 
in parameter space, which is what the Hamilton--Jacobi equation ought to be 
determining. To summarize, solving the flow equations Eqs.~(\ref{e:flow}) has 
nothing to do with solving the inflationary equations of motion.

Why is it, then, that the flow equations do seem to correspond to inflationary 
models (e.g.~Refs.~\cite{EK,peiris,KKMR})? The reason is that the ultimate 
output of the flow equations is  a function $\epsilon(\phi)$ (the evolution of 
any other slow-roll parameter could be derived from this). Long ago, I wrote a 
paper \cite{L94} which introduced the idea that a general slow-roll inflation 
model could be specified by giving the function $\epsilon(\phi)$, which 
should be less than unity for inflation to take place. This is in contrast to 
the more traditional view that a model is specified by $V(\phi)$, or perhaps 
$H(\phi)$ \cite{SB,Lidsey}, but there is a mapping between them: given 
$\epsilon(\phi)$ we can then obtain
\begin{equation}
\label{e:Hphi}
H(\phi) = H_{{\rm i}} \exp \left( \int_{\phi_{i}}^\phi \sqrt{4\pi 
\epsilon(\phi)} \, \frac{d\phi}{m_{{\rm Pl}}} \right) \,,
\end{equation}
from Eq.~(\ref{e:eps}) and 
\begin{equation}
\label{e:Vphi}
V(\phi) = \frac{3 m_{{\rm Pl}}^2}{8\pi} \, H^2(\phi) \left[ 1 - \frac{1}{3} 
\epsilon(\phi) \right] \,,
\end{equation}
from Eq.~(\ref{e:HJ}). Accordingly, for any function $\epsilon(\phi)$, which 
must be between zero and 
one, one can obtain a slow-roll inflation model with potential $V(\phi)$ which 
gives that $\epsilon(\phi)$. Indeed, Easther and Kinney \cite{EK} used a version 
of these relations to numerically obtain potentials from the flow equations.

In this light, we therefore see that what the flow equations represent is simply 
a rather complicated algorithm for generating functions $\epsilon(\phi)$, which 
have the correct general form to be interpretted as inflationary models. In 
themselves, they do not incorporate the inflationary dynamics.\footnote{This 
conclusion is somewhat less apparent for the flow equations written in terms of 
observables by Hoffman and Turner \cite{HT}, but Kinney \cite{K} showed that 
their equations are identical to the flow equations he discussed.}

That the flow equations do not incorporate the inflationary dynamics directly 
raises an interesting point --- that the outcome of the flow equation analysis 
for observable quantities would be largely unchanged even if the dynamical 
equations were different. For example, one might consider the possibility that 
the correct dynamical equations are those of the simplest braneworld model, 
based on the Type II Randall--Sundrum model \cite{RSII}, where the Friedmann 
equation is
\begin{equation}
H^2 = \frac{8\pi}{3 m_{{\rm Pl}}^2} \rho \left(1 + \frac{\rho}{2\lambda} \right) 
\,,
\end{equation}
where $\lambda$ is the brane tension. There would be a slight change to the flow 
equations, as the equation relating $N$ and $\phi$ is now changed \cite{MWBH}; 
this does not change the trajectories themselves, but it does change the measure 
of length along them, and in particular would alter the point corresponding to 
60 $e$-foldings. However the fundamental structure of the flow equations, and 
the trajectories corresponding to their solutions, are unchanged despite the 
change in the underlying dynamical assumption.

\section{Analytic solution of the flow equations}

Having concluded that the flow equations represent an algorithm for generating 
suitable functions $\epsilon(\phi)$, it is possible to discover the models that 
this procedure corresponds to, where from now on I will consider only the 
standard cosmology 
of Eq.~(\ref{e:HJ}). 
As far as I can judge, the papers in the literature so far solve the truncated 
flow equations numerically \cite{HT,K,EK}, but in fact it is possible to solve 
them analytically as follows.

The truncation requires that $^{M+1}\lambda_{{\rm H}}$ equals zero at all times. 
However if we look at the definition of that parameter, Eq.~(\ref{e:srdef}), we 
see that this is equivalent to the statement that
\begin{equation}
\frac{d^{M+2}H}{d\phi^{M+2}} = 0 \,,
\end{equation}
for all $\phi$. Hence $H(\phi)$ is a polynomial of order $M+1$, which we can 
write as
\begin{equation}
\label{e:Hsol}
H(\phi)=H_0 \left(1+A_1 \phi + \cdots + A_{M+1} \phi^{M+1} \right)
\end{equation}
where the $A_i$ are constants. Further, without loss of generality we can choose 
the initial value of $\phi$ (from which the flow equation integration begins) to 
be equal to zero.

Then the constants in Eq.~(\ref{e:Hsol}) can be easily related to the initial 
values of the slow-roll parameters for the flow equation integration, for 
example
\begin{equation}
\epsilon(\phi) = \frac{m_{{\rm Pl}}^2}{4\pi} \left( \frac{A_1 + \cdots + 
(M+1)A_{M+1} \phi^{M}}{1+A_1\phi + \cdots + A_{M+1}\phi^{M+1}} \right)^2
\end{equation}
and so 
\begin{equation}
\epsilon(\phi=0) = \frac{A_1^2 m_{{\rm Pl}}^2}{4\pi} \,,
\end{equation}
with the sign of $A_1$ determining which direction the field rolls. Similarly, 
the initial values of $^\ell\lambda_{{\rm H}}$ can be related to $A_{\ell+1}$. 
The constant $H_0$ is not fixed by the dynamical equations, but potentially can 
be determined observationally by observing the amplitude of tensor 
perturbations.

Having this closed form solution for $H(\phi)$, we can use Eq.~(\ref{e:Vphi}) to 
determine the equivalent potential, which is
\begin{eqnarray}
\label{e:Vsol}
V(\phi) &=& \frac{3 m_{{\rm Pl}}^2}{8\pi} \, H_0^2 \left(1+A_1 \phi + \cdots + 
A_{M+1} \phi^{M+1} \right)^2   \\
 && \hspace*{-26pt} \times\left[ 1 - \frac{1}{3} \frac{m_{{\rm Pl}}^2}{4\pi} 
\left( \frac{A_1 + \cdots + (M+1)A_{M+1} \phi^{M}}{1+A_1\phi + \cdots + 
A_{M+1}\phi^{M+1}} \right)^2 \right] \,.\nonumber
\end{eqnarray}
As there are so many undetermined constants, this potential can represent a wide 
range of possible behaviours. However these can be divided into two main 
classes; either the potential becomes negative, in which case slow-roll will 
fail and inflation end sometime before the potential reaches zero, or the 
potential develops a minimum at a positive value, in which case the field 
asymptotes there driving eternal inflation. These two possibilities represent 
the two late-time behaviours seen in solutions of the flow equations, with this 
attractor structure holding even if the flow equation hierarchy is taken to 
infinite order \cite{K}.

In conclusion, the flow equation approach to inflation is equivalent to 
considering the set of models described by Eq.~(\ref{e:Vsol}), where the 
constant $A_1$ is to be chosen to be consistent with the condition that 
inflation is occurring initially, i.e.~$|A_1| m_{{\rm Pl}} < \sqrt{4\pi}$, and 
the other constants are ordinarily to be chosen to satisfy the slow-roll 
conditions $|^\ell\lambda_{{\rm H}}| \ll 1$.

\section{Alternative approaches to stochastic model-building}

The previous discussion indicates that one need not employ the flow equation 
formalism in order to build up a stochastic set of inflation models by randomly 
drawing slow-roll parameters. Several approaches suggest themselves.

Firstly, the results from the flow equations approach can be reproduced by 
working directly with the parametrized potential of Eq.~(\ref{e:Vsol}). Models 
which are able to match present observations (see e.g.~Refs.~\cite{peiris,LL}) 
lie close to the extreme slow-roll limit, so it is probably sufficient to 
analyze them using the slow-roll approximation, though by construction they have 
an exact analytical solution to the equations of motion given by 
Eq.~(\ref{e:Hsol}) and so this is unnecessary.

More generally, one can ask whether there is any real benefit in using the flow 
equations to generate the function $\epsilon(\phi)$, given that they have 
nothing to do with inflationary dynamics. Working with Eq.~(\ref{e:Vsol}) is 
therefore unlikely to be any more reasonable than using simply Taylor expanding 
$V(\phi)$ itself, and then solving either using slow-roll or numerically.

As we have seen, the flow equations approach is in fact equivalent to Taylor 
expanding $H(\phi)$, Eq.~(\ref{e:Hsol}). However yet another alternative, 
perhaps the most appealing of all, is to use an expansion to generate 
$\epsilon(\phi)$ as the fundamental quantity, but to do so in a more direct way 
than the flow equations algorithm. Because $\epsilon(\phi)$ directly tells 
us whether inflation is indeed taking place, it seems an attractive starting 
point, though the need to keep it positive does restrict the allowed 
expansions. While this paper aims to elucidate the 
nature of the flow equations approach, it would be interesting to make a direct 
comparison of some of the methods outlined in this section to contrast with the 
flow equations. That ought to give some indication of whether the `preference' 
of flow-equation models for certain regions of observable parameter space 
\cite{HT,K} is a 
robust prediction, or a hidden consequence of the particular method.

\section{Conclusions}

In this paper I have shown that the flow equations approach does not directly 
incorporate inflationary dynamics. Rather, it represents a complicated algorithm 
for generating functions $\epsilon(\phi)$, which can then be used to generate 
inflationary models with dynamics matching those of the flow equations. I have 
shown that one can analytically determine the set of inflationary potentials 
which correspond to solutions of the truncated flow equations. The generality 
with which the flow equations treat inflationary dynamics therefore depends on 
the extent to which the family of potentials given by Eq.~(\ref{e:Vsol}) may 
have enough free 
parameters to be able to represent broad classes of possible potential shapes. 
However, the fact that the flow equations are so loosely related to the 
inflationary dynamics must cast some doubt on conclusions drawn from them on how 
densely inflation models sample different regions of observable parameter space.

%%%%%%%%%%%%%%%%%%%%%%%%%%%%%%%%%%%%%%%%%%%%%%%%%%%%%%%%%%%%%%%%%%%%%%%%
\begin{acknowledgments}
This work was supported in part by the Leverhulme Trust. I thank Sam Leach and 
Michael Malquarti for useful discussions and comments.
\end{acknowledgments}

%%%%%%%%%%%%%%%%%%%%%%%%%%%%%%%%%%%%%%%%%%%%%%%%%%%%%%%%%%%%%%%%%%%%%%%%
 
%%%%%%%%%%%%%%%%%%%%%%%%%%%%%%%%%%%%%%%%%%%%%%%%%%%%%%%%%%%%%%%%%%%%%%%
\end{document}